\documentclass[prd,aps,floatfix,onecolumn,amsmath,amssymb,superscriptaddress,preprintnumbers]{revtex4-2}

\usepackage{graphicx}
\usepackage{dcolumn}
\usepackage{tikz}
\usepackage{hyperref}
\usepackage{caption,subfigure,float}
\usepackage{booktabs}
\usepackage{mathrsfs}

\begin{document}

\preprint{Q.~Peng \& B.-Q.~Ma, \href{https://doi.org/10.1103/PhysRevC.107.055202}{Phys.Rev.C 107 (2022) 055202}}

\title{Determination of meson fragmentation functions in the Field-Feynman model}

\author{Qiaomu Peng}
\affiliation{\PKU}
\author{Bo-Qiang Ma}
\email{mabq@pku.edu.cn} \affiliation{\PKU}\affiliation{\CHEP}\affiliation{\CICQM}

\newcommand*{\PKU}{School of Physics, Peking University, Beijing 100871, China}

\newcommand*{\CHEP}{Center for High Energy Physics, Peking University, Beijing 100871, China}
\newcommand*{\CICQM}{Collaborative Innovation Center of Quantum Matter, Beijing, China}


\begin{abstract}
We study the fragmentation functions of both pions and kaons in the Field-Feynman recursive model with the extended SU(2) flavor symmetry relations of fragmentation functions and fitting parameters. Parametrizations are determined from a leading-order (LO) analysis of HERMES experimental multiplicity data of meson production in semi-inclusive deep inelastic scattering, and uncertainties are estimated with the Hessian method. We compare our results with the experimental data and the analysis results of other parametrizations. The SU(2) flavor symmetry breaking effect of meson fragmentation functions of $ud$ quarks is also discussed, and we show that the fragmentation functions of kaons have a bigger SU(2) flavor symmetry breaking effect of $ud$ quarks than these of pions.
\end{abstract}

\maketitle

\section{Introduction}

 Hadronization is an important topic in the field of high energy physics. Because hadronization is a non-perturbative process, it is necessary to introduce the parton fragmentation functions to study how quarks and gluons hadronize. In recent years, due to the progress of experiments, the increasing data provide new opportunity for more precise determination of fragmentation functions. At present, there already have been many parametrizations of the parton fragmentation functions such as the analysis of the de Florian, Sassot, Stratmann group (DSS)~\cite{DSS07:2007aj,DSS14:2014xna,DSS17:2017lwf} and the analysis of the Hirai, Kumano, Nagai, Sudoh group (HKNS)~\cite{Hirai:2007cx}, as well as schemes using machine learning method such as NNFF~\cite{NNFF_Bertone:2017tyb} and MAPFF~\cite{MAPFF_Khalek:2021gxf}. Although most fragmentation functions are fitted based on functional forms or neural networks, the methods based on phenomenological models are still instructive and helpful. Therefore, it is meaningful to seek for some phenomenological but intuitive models to help us understand the hadronization process, and even rely on these models to obtain the fragmentation functions of mesons.\par

In the 1970s, Field and Feynman~\cite{Field:1977fa} proposed a model based on the recursive principle to study the fragmentation process of quarks to mesons. The model has a clear and simple image, and can be simulated by the Monte Carlo method. It also inspired some further models, such as the Lund model~\cite{lund_model_Andersson:1997xwk}. The Field-Feynman model allows one to parameterize meson fragmentation functions in terms of only two parameters. Considering the flavor structure of incident quarks, Hua and Ma~\cite{Hua:2003ie} analysed the fragmentation functions in the Field-Feynman model with a significant modification. By adjusting the parameters from comparing with the parametrized fragmentation functions by Kretzer, Leader, Christova (KLC)~\cite{Kretzer:2001pz}, the fragmentation functions of the pion with different $D_u^{\pi^+},\ D_d^{\pi^+}$ and $D_s^{\pi^+}$ were obtained~\cite{Hua:2003ie} with compatible results in comparison with other empirical studies. However, due to the lack of experimental data of the kaon at that time, that work could only extend the whole framework to the kaon based on the parameters of fragmentation functions of the pion and predict the fragmentation functions of the kaon case with different $D_u^{K^+},\ D_d^{K^+},\ D_s^{K^+}$ and $D_{\bar{s}}^{K^+}$. \par

After HERMES collaboration~\cite{HERMES:2012uyd} released the final semi-inclusive deep inelastic scattering (SIDIS) data, it is time to complete the unfinished previous work and get the fragmentation functions of the kaon independently. The SIDIS process plays an important role in the determination of the flavor-separated fragmentation functions. This is because, unlike the equal amount of positive and negative charge hadrons generated in the semi-inclusive annihilation (SIA) process, the multiplicity is sensitive to the produced hadron charge and the choice of the target hadron in deep inelastic scattering (DIS). For proton targets, the multiplicity of $\pi^+$ is greater than $\pi^-$, since there are more $u$ quarks than $d$ quarks in the proton. \par

This paper is organized as follows. In Sec.~\ref{review} we briefly introduce the main aspects of the Field-Feynman model and our extension based on the previous work~\cite{Hua:2003ie}, and give the parametrized forms of the fragmentation functions of both pions and kaons. In Sec.~\ref{ana_exp}, the experimental data are analysed and the fitting procedure is discussed, including the observables in SIDIS, the selection of data sets, the estimation of uncertainties and the principle of the Hessian method. In Sec.~\ref{results}, the analysed results of the fragmentation functions of both pions and kaons are shown and compared with the experimental data and the results of other parametrizations, with the estimated uncertainties. Finally, we briefly summarize our main results in Sec.~\ref{summary}.

\section{Brief review of the Field-Feynman model}\label{review}
\begin{figure}[h]
    \centering
    \includegraphics[]{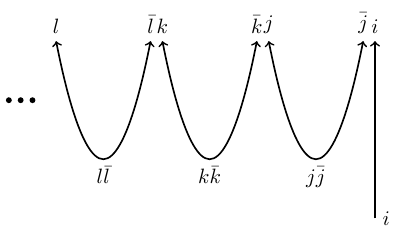}
    \caption{Illustration of the Field-Feynman model.}
    \label{fig:FFmodel}
\end{figure}

The Field-Feynman model~\cite{Field:1977fa} is a concise and intuitive phenomenological model based on the recursive principle and scaling invariance. It is illustrated in Fig.~\ref{fig:FFmodel}. An initial quark $i$ creates a color field in which new quark-antiquark pairs such as $j\bar{j}$, $k\bar{k}$, $l\bar{l},\dots$ are produced. Then quark $i$ combines with $\bar{j}$ to form a first rank primary meson and leaves $j$ to combine further antiquarks. The process goes on like a chain and produce a cascade of mesons with higher ranks.\par
To describe this process, Field and Feynman introduced a distribution function $f(\eta)$, which denotes the probability that the first rank primary meson leaves a fractional momentum $\eta$ to the remaining cascade and is normalized so that
\begin{equation}
    \int_0^1 f(\eta)\mathrm{d}\eta=1.
\end{equation}
Next they defined another distribution function $F(z)$, which means the probability of finding a meson (independent of rank) with fractional momentum $z$ in a quark jet. Then it can be written as such an integral equation

\begin{equation}
    F(z) = f(1-z) +\int_z^1 \frac{\mathrm{d}\eta}{\eta}f(\eta)F(\frac{z}{\eta}).\label{rec_equ}
\end{equation}
The first term represents the first rank primary meson, and the second term means the probability of the recursive production of a higher rank meson. Field and Feynman solved this integral equation and gave a simple solution
\begin{equation}
    zF(z)=f(1-z),
\end{equation}
where
\begin{equation}
    f(z)=(d+1)z^d.
\end{equation}
Let us define $\beta_{ij}$ as the probability of quark $i$ exciting a pair $q_j \bar{q}_j$ in quark sea, with the matrix form
\begin{equation}
    \boldsymbol\beta=
    \left(\begin{array}{ccc}
         \beta_{uu}&\beta_{ud}&\beta_{us}  \\
         \beta_{du}&\beta_{dd}&\beta_{ds}  \\
         \beta_{su}&\beta_{sd}&\beta_{ss}  \\
    \end{array}\right),
\end{equation}
and the normalization condition can be written as
\begin{equation}
    \sum_{j=1}^{n_f}\beta_{ij}=1.
\end{equation}
Here we follow the modification of the previous work by Hua and Ma~\cite{Hua:2003ie}. The original Field-Feynman model has no $i$ index but they added it and got the result of
\begin{equation}
    \boldsymbol\beta=
    \left(\begin{array}{ccc}
         0.46 & 0.46 & 0.08 \\
         0.46 & 0.46 & 0.08 \\
         0.38 & 0.38 & 0.24 \\
    \end{array}\right), \label{beta_matrix_hua}
\end{equation}
and this matrix is general for both the pion and the kaon. Obviously, there exists SU(2) flavor symmetry of $ud$ quarks between the parameters, which means that $\beta_{iu}=\beta_{id}$ and $\beta_{uj}=\beta_{dj}$, and now we abandon such relations at least for the favored quark fragmentation functions in this work.
For different quarks, the distribution $f(z)$ and $F(z)$ are also different:
\begin{align}
    f_q(z)&=(d_q+1)z^{d_q},\\
    zF_q(z)&=f_q(1-z).
\end{align}
Hence, in analogy with Eq.~\eqref{rec_equ}, we can get the mean number of fragmented mesons composed by $a\bar{b}$ with an initial quark $q$ and momentum fraction $z$:
\begin{equation}
    P_q^{a\bar{b}}(z) = \delta_{aq}\beta_{qb}f_q(1-z) + \int_z^1\frac{\mathrm{d}\eta}{\eta}f_q(\eta)\beta_{qc}P_c^{a\bar{b}}(z/\eta).\label{qab_equ}
\end{equation}
The mean number of mesons of all quark states is
\begin{equation}
    P_{\langle q\rangle}^{a\bar{b}}(z)=\sum_{c}\beta_{qc}P_{c}^{a\bar{b}}(z),
\end{equation}
then we get
\begin{equation}
    P_{\langle q\rangle}^{a\bar{b}}(z)=\beta_{qa}\beta_{qb}f_q(1-z)+\int_z^1\frac{\mathrm{d\eta}}{\eta}f_q(\eta)\beta_{qc}P_{\langle c\rangle}^{a\bar{b}}(z/\eta).\label{mean_equ}
\end{equation}
Comparing with Eq.~\eqref{rec_equ}, we can see that
\begin{equation}
    P_{\langle q\rangle}^{a\bar{b}}(z)=\beta_{qa}\beta_{qb}F_q(z).
\end{equation}
We can substitute it in Eq.~\eqref{mean_equ}, which yields
\begin{equation}
    P_{q}^{a\bar{b}}(z)=\delta_{qa}\beta_{qb}f_q(1-z) + \beta_{qa}\beta_{qb}\overline{F}_q(z),
\end{equation}
where
\begin{align}
    \overline{F}_{q}(z) &= F_q(z)-f_q(1-z)\nonumber\\
    &=(\frac{1}{z}-1)f_q(1-z)\nonumber\\
    &=(d_q+1)z^{-1}(1-z)^{d_q+1}.
\end{align}
The fragmentation function is
\begin{equation}
    D_{q}^{h}(z) = \sum_{ab}\Gamma_{a\bar{b}}^h P_{q}^{a\bar{b}}(z),
\end{equation}
where $\Gamma_{a\bar{b}}^h$ represents the probability of a meson composed by $a\bar{b}$, such as, $\Gamma_{u\bar{d}}^{\pi^+}=1$ and $\Gamma_{u\bar{u}}^{\pi^0}=\Gamma_{d\bar{d}}^{\pi^0}=
{1}/{2}$. Combining Eq.~\eqref{qab_equ} and Eq.~\eqref{mean_equ}, we get
\begin{equation}
    D_{q}^h(z)=A_q^h f_q^h(1-z)+B_q^h\overline{F}_q^h(z), \label{mainFF}
\end{equation}
where
\begin{align}
    A_q^h &= \sum_b\Gamma_{q\bar{b}}^h\beta_{qb}^h, \\
    B_q^h &= \sum_{ab}\beta_{qa}^h\Gamma^h_{a\bar{b}}\beta_{qb}^h.
\end{align}
Here we add a superscript $h$ to $\beta_{ij}$, which means the probability of a quark (an anti-quark) with flavor $i$ exciting a pair $q_j \bar{q}_j$ in quark sea during the generation of hadron $h$. In other words, it becomes a conditional probability whose condition is the generation of hadron $h$. Due to the charge conjugation, we have $\beta^{h^+}_{ij} = \beta^{h^-}_{ij}=\beta^h_{ij}$.\par

Then the parametrized form of $\pi^+$ fragmentation functions can be written as
\begin{align}
    D_{u}^{\pi^+}(z) &=\beta_{ud}^{\pi} f_u^{\pi}(1-z)+\beta_{uu}^{\pi}\beta_{ud}^{\pi} \overline{F}_u^{\pi}(z), \label{Dpiu}\\
	D_{\bar{d}}^{\pi^+}(z) &=\beta_{du}^{\pi} f_d^{\pi}(1-z)+\beta_{du}^{\pi}\beta_{dd}^{\pi} \overline{F}_d^{\pi}(z), \label{Dpidbar}\\
	D_{d}^{\pi^+}(z) &=\beta_{du}^{\pi}\beta_{dd}^{\pi} \overline{F}_d^{\pi}(z), \\
	D_{\bar{u}}^{\pi^+}(z) &=\beta_{uu}^{\pi}\beta_{ud}^{\pi} \overline{F}_u^{\pi}(z),	\\
	D_{s}^{\pi^+}(z) &=D_{\bar{s}}^{\pi^+}(z)=\beta_{su}^{\pi}\beta_{sd}^{\pi} \overline{F}_s^{\pi}(z).
\end{align}
In the same way, the fragmentation functions of $K^+$ can also be obtained:
\begin{align}
    D_{u}^{K^+}(z) &=\beta_{us}^{K} f_u^{K}(1-z)+\beta_{uu}^{K}\beta_{us}^{K} \overline{F}_u^{K}(z), \label{Dku}\\
	D_{\bar{s}}^{K^+}(z) &=\beta_{su}^{K} f_s^{K}(1-z)+\beta_{su}^{K}\beta_{ss}^{K} \overline{F}_s^{K}(z), \label{Dksbar}\\
	D_{s}^{K^+}(z) &=\beta_{su}^{K}\beta_{ss}^{K} \overline{F}_s^{K}(z), \\
	D_{\bar{u}}^{K^+}(z) &=\beta_{uu}^{K}\beta_{us}^{K} \overline{F}_u^{K}(z),	\\
	D_{d}^{K^+}(z) &=D_{\bar{d}}^{K^+}(z)=\beta_{du}^{K}\beta_{ds}^{K} \overline{F}_d^{K}(z).
\end{align}
It can be derived that the corresponding fragmentation functions of negatively charged mesons can be obtained by charge conjugation $D_q^{h^+}(z)=D_{\bar{q}}^{h^-}(z)$ in the Field-Feynman model. But for neutral mesons, the relation is not simply $D_{q}^{h^0}(z)=[D_{q}^{h^+}(z)+D_{q}^{h^-}(z)]/2$. For brevity, we put the derivations in  Appendix~\ref{appendix_a}.\par
Since the parameters $\beta_{su}^{\pi}$ and $\beta_{sd}^{\pi}$ only appear in $D_s^{\pi^+}(z)$ in the form of a product $\beta_{su}^{\pi}\beta_{sd}^{\pi}$, they can not be completely determined in the fitting procedure, so we assume that $\beta_{su}^{\pi}=\beta_{sd}^{\pi}$ to constrain them. Similarly, there exists the same problem with the parameters $\beta_{du}^{K}$ and $\beta_{ds}^{K}$ in $D_{d}^{K^+}(z)$, but the constraint becomes $\beta_{ds}^{K}=1-2\beta_{du}^{K}$ (equivalent to $\beta^{K}_{du}=\beta^{K}_{dd}$). Actually, the assumptions and constraints of these parameters in the original Field-Feynman model are exactly the same, and we just follow them here.\par

In total we have 8 free parameters to describe the fragmentation functions of each meson, which are
\begin{align*}
    &\pi:\quad d^{\pi}_u, d^{\pi}_d, d^{\pi}_s, \beta^{\pi}_{uu}, \beta^{\pi}_{ud}, \beta^{\pi}_{du}, \beta^{\pi}_{dd}, \beta^{\pi}_{su};\\
    &K:\quad d^{K}_u, d^{K}_d, d^{K}_s, \beta^{K}_{uu}, \beta^{K}_{us}, \beta^{K}_{du}, \beta^{K}_{su}, \beta^{K}_{ss}.\\
\end{align*}
In this work, the parameters $d_u\neq d_d$, which leads to the breaking of SU(2) flavor symmetry of $ud$ quarks between fragmentation functions (such as $D_u^{\pi^+}(z)\neq D_{\bar{d}}^{\pi^+}(z)$). In fact, the mass difference between $u$ and $d$ quarks could inevitably make the SU(2) flavor symmetry of $ud$ quarks no longer be maintained in the process of hadronization.

\section{Analysis of experimental data}\label{ana_exp}
\subsection{Observables in SIDIS and data selection}
In hadronization of SIDIS, the relevant observable is multiplicity for hadrons of a specific type $h$, and such observable is defined as the differential cross section for hadron production normalized to the differential inclusive DIS cross section. With leading order (LO) QCD analysis, multiplicity can be expressed as
\begin{equation}
    \frac{1}{\sigma_{\mathrm{tot}}^{\text{DIS}}}\frac{\mathrm{d}\sigma^{h}}{\mathrm{d}z} = \frac{\sum_{f}e_f^2 \int_{x_\mathrm{min}}^{x^\mathrm{max}}\mathrm{d}x_B q_f(x_B,Q^2)D_f^{h}(z,Q^2)}{\sum_f e_f^2\int_{x_\mathrm{min}}^{x^\mathrm{max}}\mathrm{d}x_B q_f(x_B,Q^2)},\label{multi_calc}
\end{equation}
where the sum is over quarks and antiquarks of flavor $f$, and $e_f$ is the charge of quark in units of elementary charge. $q_f(x_B, Q^2)$ is the quark parton distribution function (PDF) with Bjorken variable $x_B=-q^2/(2P\cdot q)$ and negative square of the four-momentum transfer $Q^2=-q^2$. $D_f^{h}(z,Q^2)$ is the fragmentation function with momentum fraction $z=(P\cdot p_h)/(P\cdot q)$. Due to the scaling invariance of the Field-Feynman model~\cite{Field:1989uq}, we have $D_f^{h}(z,Q^2)=D_f^{h}(z)$.

We choose the $z$ presentation of the final HERMES data~\cite{HERMES:2012uyd} on pion and kaon multiplicities for both proton and deuteron targets to present this LO analysis. There are a total of 8 groups of data (see Fig.~8 in Ref.~\cite{HERMES:2012uyd}) from different scattering processes, with 10 data points in each group 
fully used in this analysis. The information of data are listed in Table~\ref{hermesdata}. The determination of fragmentation functions requires knowledge of PDFs of the proton and deuteron targets, for which we use MSTW08 LO~\cite{mstw08_Martin:2009iq} parametrization scheme with the accepted integral range $0.023 < x_B < 0.600$.

\begin{table}[h]
    \centering
    \caption{HERMES~\cite{HERMES:2012uyd} multiplicity data used in our LO analysis, with the mean value of $Q^2 = 2.5\ \mathrm{GeV}^2$.}
    \begin{ruledtabular}
    \begin{tabular}{lcc}
       \makebox[1em]{} process \makebox[1em]{} & data in fit &\makebox[1em]{} $\chi^2$ per data \makebox[1em]{}\\
    \hline
        $ep\ \to \ \pi^+X$ & 10 & 1.446\\
        $ep\ \to \ \pi^-X$ & 10 & 1.843\\
        $ed\ \to \ \pi^+X$ & 10 & 1.886\\
        $ed\ \to \ \pi^-X$ & 10 & 1.240\\
        $ep\ \to \  K^+X$ & 10 & 1.168\\
        $ep\ \to \  K^-X$ & 10 & 0.988\\
        $ed\ \to \  K^+X$ & 10 & 1.486\\
        $ed\ \to \  K^-X$ & 10 & 1.034\\
    \hline
        \makebox[1em]{} total & 80 & 1.387\\
    \end{tabular}
    \end{ruledtabular}
    \label{hermesdata}
\end{table}
\subsection{Fitting procedure and uncertainty analysis}
By comparing HERMES~\cite{HERMES:2012uyd} experimental multiplicity data and the theoretical values calculated with Eq.~\eqref{multi_calc}, we determine the optimal values of the 8 independent fit parameters by minimizing the $\chi^2$ function
\begin{equation}
    \chi^2=N\sum_i\frac{(M_i^{\text{data}}-M_i^{\text{theory}})^2}{\sigma_i^2},
\end{equation}
where the sum on $i$ is over the number of data points, $N=\sum_i\sigma_i^2$ is a rescaling factor to avoid the distortion of $\chi^2$ caused by too small $\sigma_i$, and $M_i^{\text{data}}$ and $M_i^{\text{theory}}$ are experimental and theoretical values of multiplicity respectively. The experimental errors are calculated from systematic and statistical errors by $\sigma_i^2=(\sigma_i^{\text{sys}})^2+(\sigma_i^{\text{stat}})^2$, and the bins of $z$ are treated as horizontal systematic errors and calculated by effective variance method~\cite{EVmethod} with the built-in function of ROOT~\cite{root_Brun:1997pa}. Because we only select the $z$ presentation of the HERMES data, the bin correlations between $z$ and other kinematic variables are neglected. \par
In addition to accurately determining each parameter, it is also important to estimate their uncertainties. In this study, we apply the Hessian method~\cite{Pumplin:2001ct} to evaluate the uncertainty band of fragmentation functions. The Hessian method is based on a quadratic expansion of the $\chi^2$ function around its global minimum point $\hat{\alpha}$:
\begin{equation}
    \Delta\chi^2(\hat{\alpha})=\chi^2(\hat{\alpha}+\delta\alpha) - \chi^2(\hat{\alpha})=\sum_{i,j}H_{ij}\delta\alpha_i\delta\alpha_j,
\end{equation}
where the sum of $i$ and $j$ is over the 8 free fit parameters, and $\Delta\chi^2(\hat{\alpha})$ is the deviation from the minimum, $\delta\alpha_i$ are the parameter errors around the minimum.\par
The Hessian matrix is constructed by the second derivatives of $\chi^2$ function at the minimum and its definition is $H_{ij}=\frac{1}{2}\frac{\partial\chi^2}{\partial\alpha_i\partial\alpha_j}\big|_{\hat{\alpha}}$. It is a symmetric $d\times d$ matrix where $d$ is the number of free fit parameters (or degrees of freedom). The uncertainties of fragmentation functions are obtained by
\begin{equation}
    [\delta D_f^{h}(z)]^2=\Delta\chi^2\sum_{i,j}\left(\frac{\partial D_f^{h}(z,\hat{\alpha})}{\partial\alpha_i}\right)H_{ij}^{-1}\left(\frac{\partial D_f^{h}(z,\hat{\alpha})}{\partial\alpha_j}\right).\label{deltaDz}
\end{equation}
The derivation of Eq.~\eqref{deltaDz} can be found in Ref.~\cite{Pumplin:2000vx}. We set the tolerance parameter $T^2=\Delta\chi^2=9.3028$ in this fitting for $68\%$ confidence level by solving the integral equation of $\chi^2$ distribution function
\begin{equation}
    P=\int_0^{\Delta\chi^2}\frac{1}{2\Gamma(n/2)}\left(\frac{x}{2}\right)^{\frac{n-2}{2}}\exp\left(\frac{x}{2}\right)\mathrm{d} x = 0.6826,
\end{equation}
where $\Gamma(n/2)$ is the Gamma function and $n=8$ is the degree of freedom (the number of fitting parameters). This calculation method is referred to Ref.~\cite{Hirai:2007cx}. The minimization of $\chi^2$ function and calculation of Hessian matrix are evaluated by CERN subroutine ROOT~\cite{root_Brun:1997pa} and \textsc{Minuit2}~\cite{minuit2_Hatlo:2005cj}.
\section{Results}\label{results}
We get the following results. At the same time, we refit the model used in the previous work~\cite{Hua:2003ie} according to the HERMES~\cite{HERMES:2012uyd} multiplicity data for comparison.

\subsection{Optimum parameters and comparisons with experimental data}\label{subsec:pars}
\begin{figure}[h]
    \centering
    \subfigure[]{
    \begin{minipage}{0.49\linewidth}
    \includegraphics[width=0.9\textwidth]{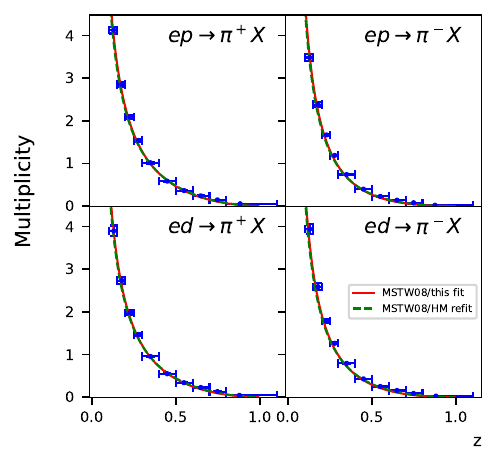}
    \end{minipage}}
    \subfigure[]{
    \begin{minipage}{0.49\linewidth}
    \includegraphics[width=0.9\textwidth]{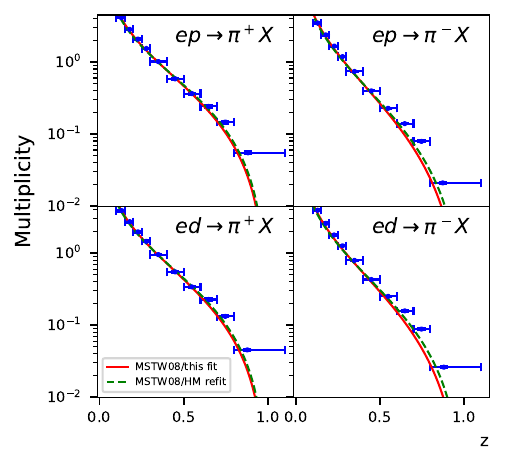}
    \end{minipage}}
    \caption{(color online) Comparison of the analysis result with HERMES pion multiplicity data, the red line is our fit results and the green dash line is refitted results of the previous work~\cite{Hua:2003ie}. (a) and (b) are identical but the $y$ axis of (b) is logarithmic. We can see that there is no significant difference between them.}
    \label{fig:mult_pion}
\end{figure}

\begin{figure}[h]
    \centering
    \subfigure[]{
    \begin{minipage}{0.49\linewidth}
    \includegraphics[width=0.9\textwidth]{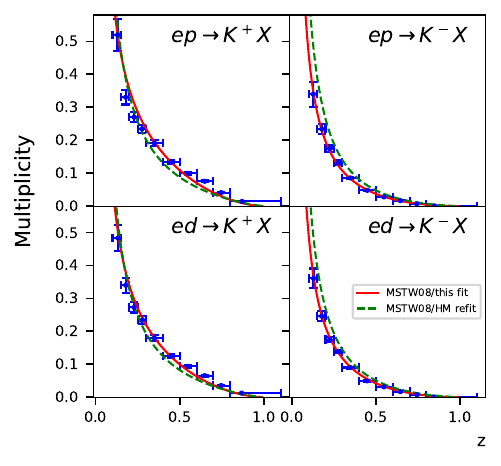}
    \end{minipage}}
    \subfigure[]{
    \begin{minipage}{0.49\linewidth}
    \includegraphics[width=0.9\textwidth]{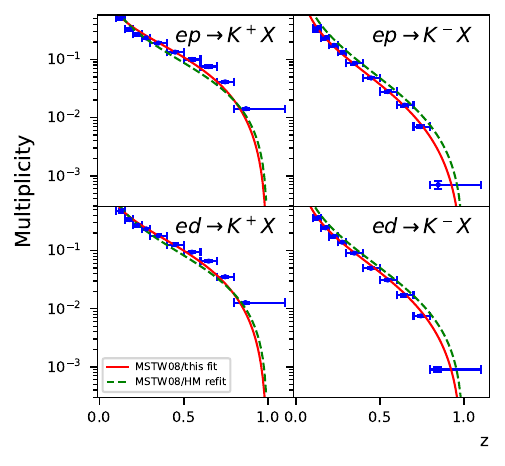}
    \end{minipage}}
    \caption{(color online) The same as in Fig.~\ref{fig:mult_pion}, but now for the kaon. We can see that there are significant differences between our fit results and refitted results, and the fitting of refitted results is not as good as our fit results.}
    \label{fig:mult_kaon}
\end{figure}

Obtained parameters in the LO analysis are listed in Tables~\ref{tab:pars_pion} and \ref{tab:pars_kaon} for both the pion and the kaon. In Fig.~\ref{fig:mult_pion} and Fig.~\ref{fig:mult_kaon} we present a detailed comparison of the results of our fit with HERMES~\cite{HERMES:2012uyd} experimental multiplicity data and the refitted result of previous work~\cite{Hua:2003ie}, respectively. In general, the agreement of the fit with multiplicity data is excellent in the entire $z$ range covered by the experiments. However, in Fig.~\ref{fig:mult_kaon} the refitted results are not in good agreement with the experimental data of the kaon, while their performance is very good in terms of the pion in Fig.~\ref{fig:mult_pion}.

\begin{table}[h]
    \centering
    \caption{Parameters determined for the pion, with the positions of parameters selected according to their corresponding relations after extension. The definitions of refitted parameters can be referred to Ref.~\cite{Hua:2003ie}}
    \begin{ruledtabular}
    \begin{tabular}{cccc}
        \multicolumn{2}{c}{this fit} & \multicolumn{2}{c}{HM refit}\\[5pt]
         parameters &  value & parameters & value\\
         \hline
         $d_u^{\pi}$ & $1.724\pm 0.523$ &$d$ & $1.653\pm 0.247$ \\
         $d_d^{\pi}$ & $2.205\pm 0.479$ &&\\
         $d_s^{\pi}$ & $1.777\pm 0.586$ & &\\
         $\beta_{uu}^{\pi}$ & $0.487\pm 0.022$ &$\beta_u$ &$0.469\pm 0.012$\\
         $\beta_{ud}^{\pi}$ & $0.470\pm 0.014$ & &\\
         $\beta_{du}^{\pi}$ & $0.491\pm 0.011$ & &\\
         $\beta_{dd}^{\pi}$ & $0.404\pm 0.041$ & &\\
         $\beta_{su}^{\pi}$ & $0.430\pm 0.018$ & $\beta_s$&$0.472\pm 0.014$\\
    \end{tabular}
    \end{ruledtabular}
    \label{tab:pars_pion}
\end{table}

\begin{table}[h]
    \centering
    \caption{Same as Table~\ref{tab:pars_pion} but for the kaon.}
    \begin{ruledtabular}
    \begin{tabular}{cccc}
         \multicolumn{2}{c}{this fit} & \multicolumn{2}{c}{HM refit}\\[5pt]
         parameters &  value & parameters & value\\
        \hline
        $d_u^{K}$ & $1.709\pm 0.307$ &$d$& $1.279\pm 0.195$\\
        $d_d^{K}$ & $2.126\pm 0.116$ & &\\
        $d_s^{K}$ & $2.757\pm 0.281$ & &\\
        $\beta_{uu}^{K}$ & $0.102\pm 0.056$ &$\beta_u$ &$0.464\pm 0.015$\\
        $\beta_{us}^{K}$ & $0.153\pm 0.131$ & &\\
        $\beta_{du}^{K}$ & $0.471\pm 0.036$ & &\\
        $\beta_{su}^{K}$ & $0.863\pm 0.016$ &$\beta_s$ &$0.481\pm 0.031$\\
        $\beta_{ss}^{K}$ & $0.050\pm 0.013$ & &\\
    \end{tabular}
    \end{ruledtabular}
    \label{tab:pars_kaon}
\end{table}


For the pion, the refitted results are almost the same as the results of our fit. This can be observed not only from Fig.~\ref{fig:mult_pion}, but also from the comparison of matrices Eq.~\eqref{beta_matrix_hua} and
\begin{equation}
    \boldsymbol{\beta}^{\pi} =
    \left(\begin{array}{ccc}
	   0.487 & 0.470 & 0.043\\
        0.491 & 0.404 & 0.105\\
        0.430 & 0.430 & 0.139
\end{array}\right).
\end{equation}
Since the difference between the two models (ours and the previous one) is whether the SU(2) flavor symmetry of $ud$ quarks is maintained, it shows that the SU(2) flavor symmetry breaking effect of $ud$ quarks is not significant for the pion.\par
\begin{figure}[h]
    \centering
    \subfigure[]{
    \begin{minipage}{0.49\linewidth}
        \includegraphics[width=0.7\textwidth]{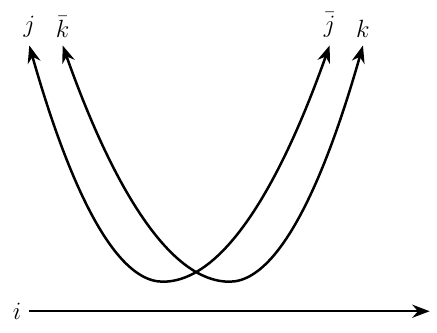}
    \end{minipage}}
    \subfigure[]{
    \begin{minipage}{0.49\linewidth}
        \includegraphics[width=0.7\textwidth]{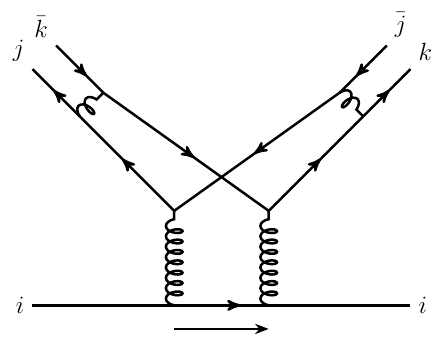}
    \end{minipage}}
    \caption{Such a process is ignored in the Field-Feynman model,  but it is possible to contribute to the generation of mesons. (a) is the illustration and (b) is the corresponding Feynman diagram. Obviously, gluons play a major role in this process.}
    \label{fig:additional_contribution}
\end{figure}

But for the kaon, the situation is different. The refitted results deviate more severely from the experimental data than our fit results for the kaon in Fig.~\ref{fig:mult_kaon}. Comparing the matrix
\begin{equation}
    \boldsymbol{\beta}^{K}=
    \left(\begin{array}{ccc}
	   0.102 & 0.745 & 0.153\\
        0.471 & 0.471 & 0.059\\
        0.863 & 0.087 & 0.050
    \end{array}\right)  \label{beta_matrix_kaon}
\end{equation}
with Eq.~\eqref{beta_matrix_hua}, we can see that some of their elements are also very different. It can be inferred that the SU(2) flavor symmetry breaking effect of the kaon fragmentation functions of $ud$ quarks is relatively large. Apparently, it is inappropriate to continue assuming SU(2) flavor symmetry of $ud$ quarks for the kaon in the model. In addition, from the matrix we can notice that the matrix element $\beta^K_{su}$ is much larger than other matrix elements in the same row (as a probability). It shows that the first rank processes, i.e., the excitation of $u\bar{u}$ from $\bar{s}$ (or $s$) and the combination of $u\bar{s}$ (or $\bar{u}s$), are dominant in the hadronization of the kaon. This is consistent with the definition of $\beta^{K}_{su}$ as a conditional probability and reflects the posterior nature given by the analysis of the experimental data of the kaon, which means when the detector observes a charged kaon event, it is
probably the result of the excitation of $u\bar{u}$ from $\bar{s}$ (or $s$) and the subsequent combination of $u\bar{s}$ (or $\bar{u}s$). There is a similar problem with $\beta^K_{ud}$, but we should not regard it in the same way. In fact, the large value of $\beta^K_{ud}$ is not a physical result. According to Eq.~\eqref{Dku}, only parameters $\beta^K_{uu}$ and $\beta^K_{us}$ can be determined in the fitting procedure, where $\beta^K_{us}$ is determined by the second term and the product of $\beta^K_{us}$ and $\beta^K_{uu}$ is determined by the first term. The parameter $\beta^{K}_{ud}$ is obtained by the relation $\beta^K_{ud}=1-\beta^K_{uu}-\beta^K_{us}$. This leads to a problem that $\beta^K_{uu}$ is not determined independently and would depend on the value of $\beta^K_{us}$. The parameter $\beta^K_{us}$ means the probability of a $u$ quark exciting an $s\bar{s}$ pair in the generation of the kaon. From the perspective of quantum field theory, because the mass of $s$ quarks is much larger than that of $u$ quarks, the probability of a $u$ quark exciting an $s\bar{s}$ pair is extremely low (approximately less than $10^{-3}$). But if we take the generation of the kaon as a condition, this probability would become quite considerable. This would depress the value of the parameter $\beta^K_{uu}$ in the fitting procedure, and leads to the high value of parameter $\beta^K_{ud}$. However, this explanation is only mathematical. Here we propose a possible physical explanation. In the meson generation process, gluons also have independent contributions, but there are no gluons in the Field-Feynman model. This leads to the contributions from gluons being counted into quarks. As shown in Fig.~\ref{fig:additional_contribution}, a process like this will produce one positive and one negative mesons with the same momentum, but the Field-Feynman model does not take such a process into consideration.
 For the pion, this neglect will not have many effects, but for the kaon, it is not the same because there is no $s$ or $\bar{s}$ quarks (valence) in protons (or deuterons) and it is difficult for sea quarks alone to provide enough $s$ or $\bar{s}$ quarks to generate kaons. There are more kaons to come from $s\bar{s}$ quark pairs produced by gluons. The gluons are mainly excited by $u$ quarks since there are two $u$ quarks (valence) in one proton (and three in one deuteron). The model does not consider the contribution of $u$ quarks through gluons, and the parameter $\beta^K_{ud}$ is not directly obtained by fitting, so this part of the contribution is considered to be generated through $d$ quarks in the calculation. Nevertheless, we still have the opportunity to analyse the contribution of gluons in the future. As long as the multiplicity data of the neutral kaons $K^{0}$ and $\bar{K}^{0}$ are obtained experimentally, the fragmentation functions of the neutral kaon in  Appendix~\ref{appendix_a} can be fitted. Then the parameter $\beta^K_{ud}$ can be determined, so that the contribution of gluons would be known after deducting the contribution of $d$ quarks.\par

\begin{figure}
    \centering
    \includegraphics[width=0.5\textwidth]{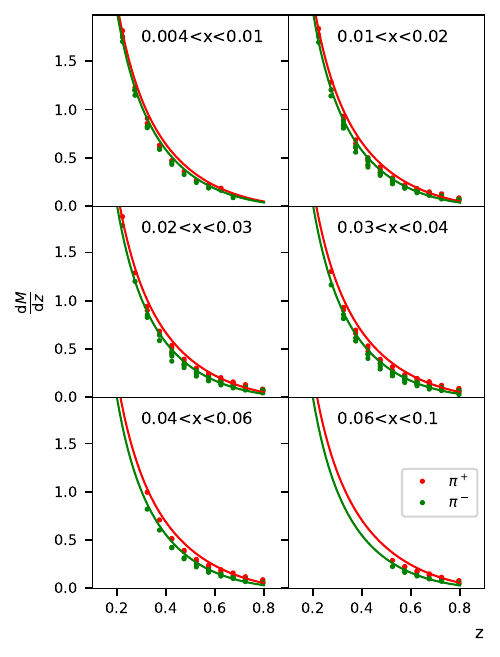}
    \caption{(color online) Comparison of the analysis result with COMPASS pion differential multiplicity data~\cite{COMPASS:2016xvm} with range $1\ \mathrm{GeV}^2<Q^2<4\ \mathrm{GeV}^2$, the red and green dots are positively and negatively charged pion data, and the red and green lines are our analysis results correspondingly (with MSTW08 LO~\cite{mstw08_Martin:2009iq} PDFs).}
    \label{fig:compass_exp_pion}
\end{figure}

\begin{figure}
    \centering
    \includegraphics[width=0.5\textwidth]{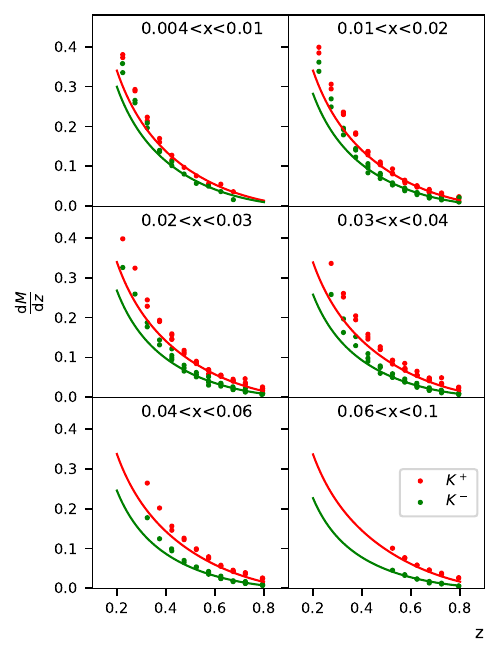}
    \caption{(color online) Same as Fig.~\ref{fig:compass_exp_pion}, but for kaons.}
    \label{fig:compass_exp_kaon}
\end{figure}

 The COMPASS experimental data~\cite{COMPASS:2016xvm,COMPASS:2016crr} can verify the validity of our fragmentation functions. The observable of COMPASS experiment is differential multiplicity. With LO QCD analysis, it can be expressed as
 \begin{equation}
     \frac{\mathrm{d}M^h(x,z,Q^2)}{\mathrm{d}z} = \frac{\sum_f e_f^2 q_f(x,Q^2) D_f^h(z,Q^2)}{\sum_f^2 e_f^2 q_f(x,Q^2)}.
 \end{equation}
 We select a part of COMPASS experimental data with range $1\ \mathrm{GeV}^2<Q^2<4\ \mathrm{GeV}^2$ (close to $2.5\ \mathrm{GeV}^2$ of HERMES). Fig.~\ref{fig:compass_exp_pion} and Fig.~\ref{fig:compass_exp_kaon} show the experimental and theoretical values of differential multiplicities of pions and kaons respectively. It can be seen that there is some inconsistency between the experimental and theoretical values. We think it is caused by the differences between the two experiments. In Refs.~\cite{COMPASS:2016xvm,COMPASS:2016crr}, the sum of integrated meson multiplicities $\mathscr{M}^{h^+}+\mathscr{M}^{h^-}$ is introduced to compare the data, with $\mathscr{M}^{h^{\pm}} = \int\langle M^{h^\pm}(x,y,z)\rangle_y\mathrm{d}z$. In this paper, for pions, the experimental values are lower than the theoretical values. This is consistent with the conclusion in Ref.~\cite{COMPASS:2016xvm}. The sum of integrated pion multiplicities of COMPASS is lower than that of HERMES. For kaons, the experimental values are higher than the theoretical values, which is consistent with the conclusion in Ref.~\cite{COMPASS:2016crr}. The sum of integrated kaon multiplicity of COMPASS is higher than HERMES. From comparing the conditions of COMPASS and HERMES, the inconsistency is mainly caused by different $\sqrt{s}$ of the two experiments.

It should be noted that the fit parameters $\beta^{\pi}_{su}$ and $\beta^{\pi}_{sd}$ could not be determined in the fitting procedure since they appear in the fragmentation functions of the pion as a product
$\beta^{\pi}_{su}\beta^{\pi}_{sd}$
($\beta^{K}_{du}$ and $\beta^{K}_{ds}$ are the same). We have adopted a compromise approach, that is, assuming that they are still in accordance with the relation $\beta^{\pi}_{su}=\beta^{\pi}_{sd}$ in the original Field-Feynman model (for the kaon which is $\beta^{K}_{ds}=1-2\beta^{K}_{du}$), but this may not be the realistic case. This makes these parameters have greater uncertainties and a discussion for the value ranges of these parameters is given in Appendix~\ref{appendix_b}.

\subsection{Uncertainties and comparison with other parametrizations}

\begin{figure}[h]
    \centering
    \includegraphics[width=0.8\textwidth]{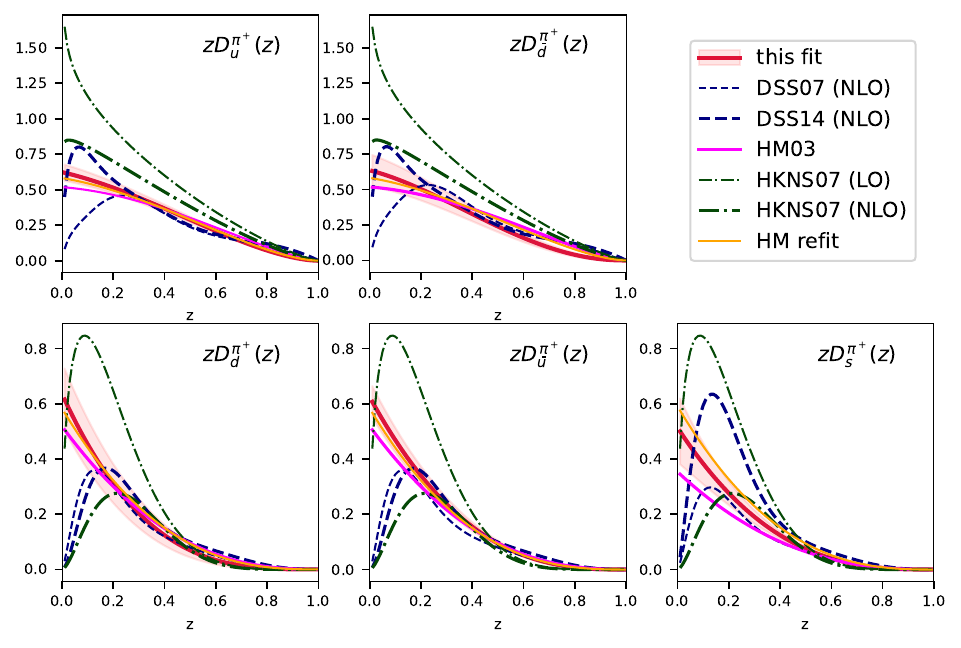}
    \caption{(color online) The individual fragmentation functions for the positively charged pion $zD_i^{\pi^+}(z)$ along with uncertainty indicated by the shaded bands. Also shown is a comparison to the refitted results, the previous analysis of HM~\cite{Hua:2003ie} and other parametrizations by DSS~\cite{DSS07:2007aj, DSS14:2014xna} and HKNS~\cite{Hirai:2007cx}.}
    \label{my_pion_errorband}
\end{figure}

\begin{figure}[h]
    \centering
    \includegraphics[width=0.8\textwidth]{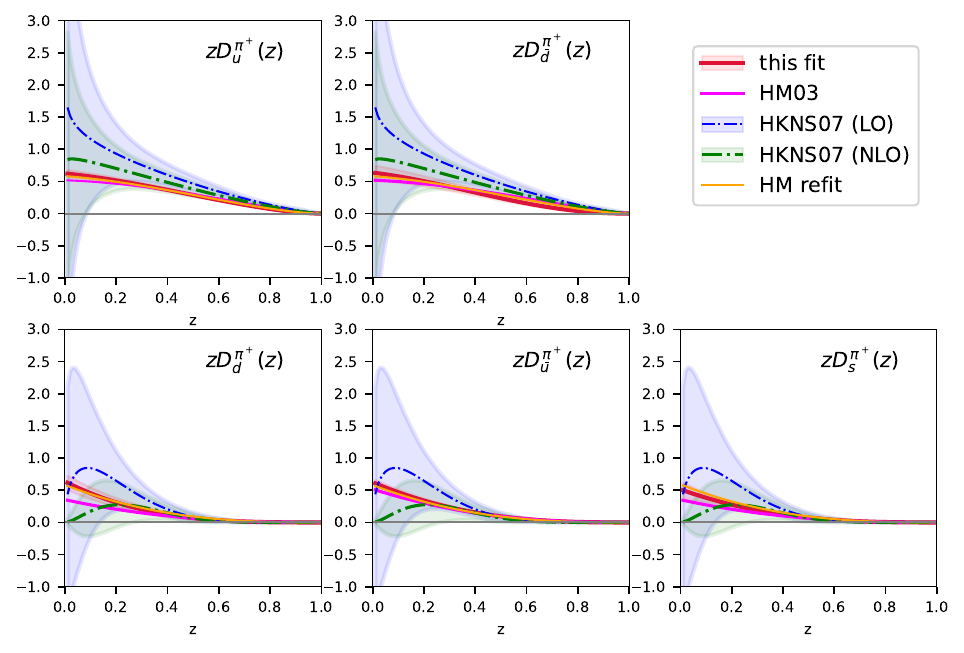}
    \caption{(color online) The same as in Fig.~\ref{my_pion_errorband} without the parametrization by DSS~\cite{DSS07:2007aj, DSS14:2014xna}, and the uncertainty of HKNS~\cite{Hirai:2007cx} is shown with shaded bands. }
    \label{hkns_pion_errorband}
\end{figure}

\begin{figure}[h]
    \centering
    \includegraphics[width=0.8\textwidth]{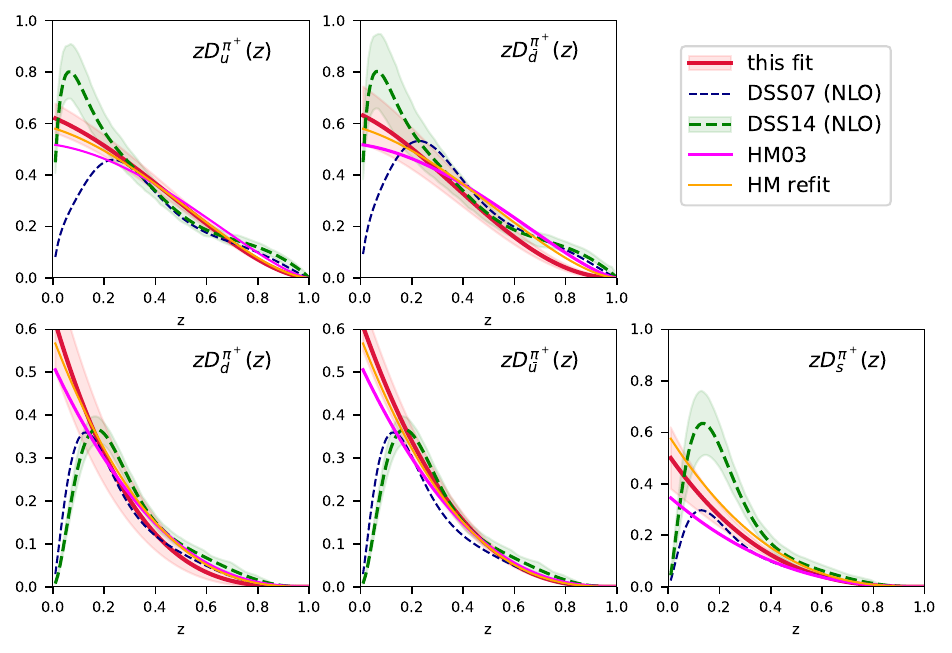}
    \caption{(color online) The same as in Fig.~\ref{my_pion_errorband} without the parametrization by HKNS~\cite{Hirai:2007cx}, and the uncertainty of DSS~\cite{DSS14:2014xna} is shown with shaded bands.}
    \label{dss_pion_errorband}
\end{figure}

The obtained fragmentation functions and their uncertainties are shown for the positively charged pion in Fig.~\ref{my_pion_errorband}. The shaded areas indicate their one-$\sigma$ uncertainty (68\% confidence level) regions estimated by the Hessian method. The comparison with other parameterizations is also shown in the same figure. In order to avoid the confusion caused by the superposition of more than three kinds of shaded areas, their uncertainties are shown in Fig.~\ref{hkns_pion_errorband} and Fig.~\ref{dss_pion_errorband} respectively. \par

Since the functions we adopt are monotonically decreasing and the forms are relatively simple, the behaviors of fragmentation functions of other parametrizations are more complex than ours, especially in the region $0<z<0.2$. But in the region $0.2<z<0.6$, the trends of our fragmentation functions are consistent with other parametrizations. There are no peaks for either favored or unfavored fragmentation functions, according to Eq.~\eqref{Dpiu} and Eq.~\eqref{Dpidbar}, which means that the contribution of higher rank primary meson accounts more than first rank primary meson in the generation of the pion.\par
\begin{figure}[h]
    \centering
    \includegraphics[width=0.8\textwidth]{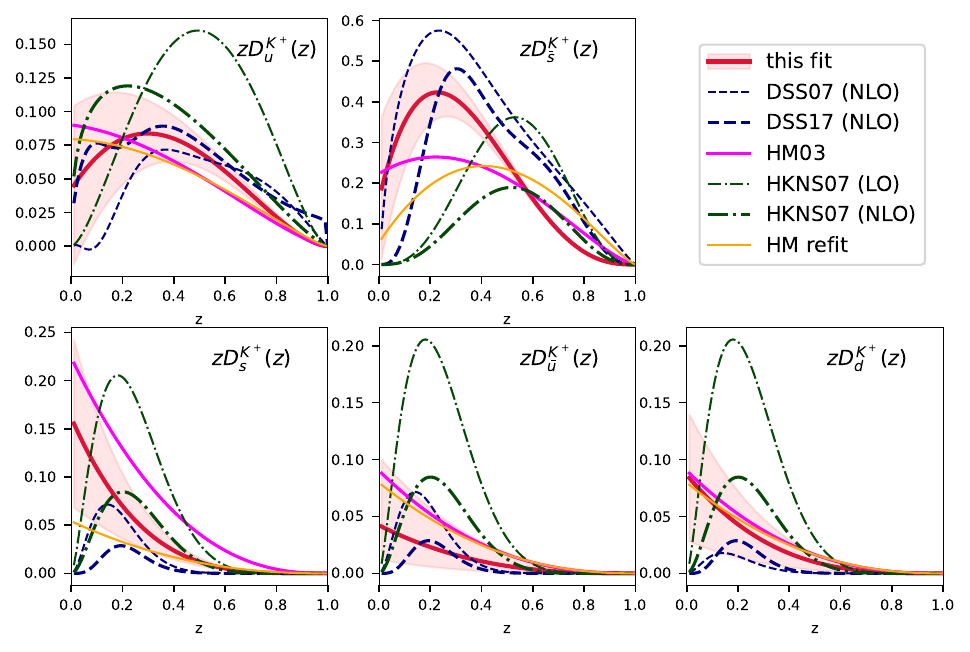}
    \caption{(color online) The individual fragmentation functions for the positively charged kaon $zD_i^{K^+}(z)$ along with uncertainty indicated by the shaded bands. Also shown is a comparison to the refitted results, the previous analysis of HM~\cite{Hua:2003ie} and other parametrizations by DSS~\cite{DSS07:2007aj, DSS17:2017lwf} and HKNS~\cite{Hirai:2007cx}.}
    \label{my_kaon_errorband}
\end{figure}

\begin{figure}[h]
    \centering
    \includegraphics[width=0.8\textwidth]{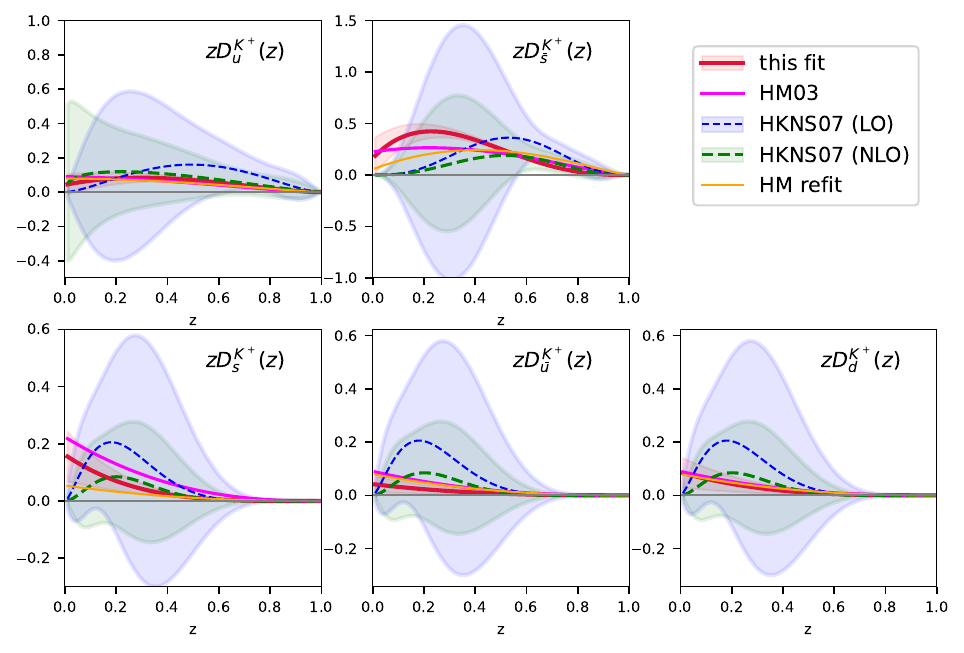}
    \caption{(color online) The same as in Fig.~\ref{my_kaon_errorband} without the parametrization by DSS~\cite{DSS07:2007aj, DSS17:2017lwf}, and the uncertainty of HKNS~\cite{Hirai:2007cx} is shown with shaded bands. }
    \label{hkns_kaon_errorband}
\end{figure}

\begin{figure}[h]
    \centering
    \includegraphics[width=0.8\textwidth]{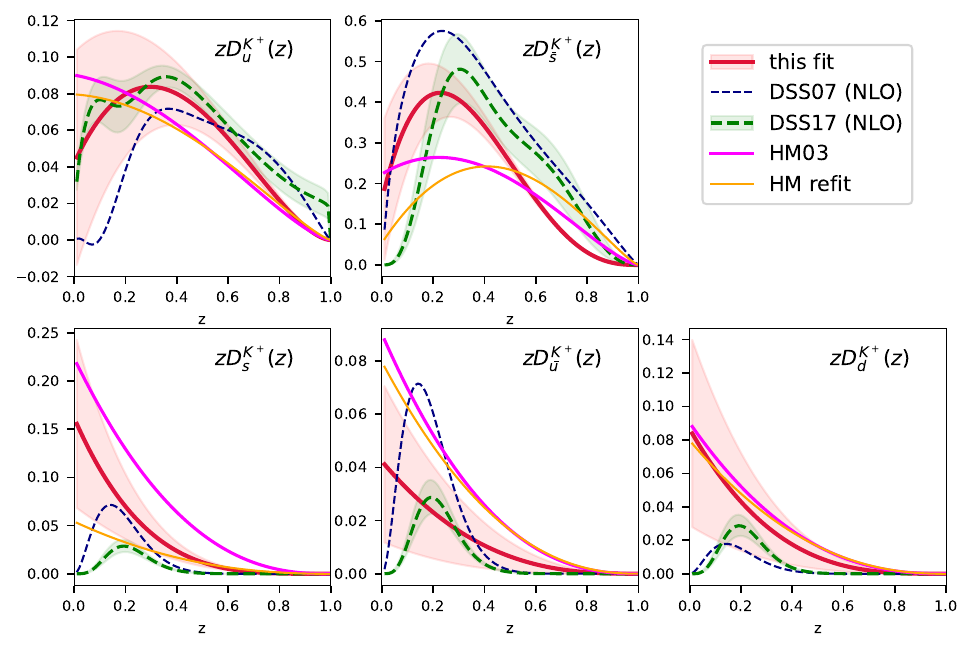}
    \caption{(color online) The same as in Fig.~\ref{my_kaon_errorband} without the parametrization by HKNS~\cite{Hirai:2007cx}, and the uncertainty of DSS~\cite{DSS17:2017lwf} is shown with shaded bands.}
    \label{dss_kaon_errorband}
\end{figure}

The obtained fragmentation functions and their uncertainties are shown for the positively charged kaon in Fig.~\ref{my_kaon_errorband}. To avoid confusion caused by superposition of shaded area, uncertainties of other parametrizations are shown in Figs.~\ref{hkns_kaon_errorband} and \ref{dss_kaon_errorband}. Different from the pion results, the fragmentation functions of the kaon show relatively greater uncertainty and more complex behaviors, especially for the favored functions $D_u^{K^+}(z)$ and $D_{\bar{s}}^{K^+}(z)$. We can see that they have peaks in the region $0.2<z<0.4$. According to Eq.\eqref{Dku} and Eq.\eqref{Dksbar}, it shows that the contribution of the first rank primary mesons accounts for a large proportion in the generation of the kaon, which is consistent with the analysis of the parameters of the kaon in Sec.~\ref{subsec:pars}.

There are significant discrepancies between our fragmentation functions and DSS schemes in shape and uncertainty especially for the kaon. We guess this is caused by two reasons. On the one hand, the fragmentation function form of DSS is $D(z)\propto Nz^\alpha(1-z)^\beta[1+\gamma (1-z)^\delta]$, which has more free parameters. On the other hand, DSS fit has adopted a wider range of experimental data sources (include both SIA and SIDIS).

\section{Summary and conclusion}\label{summary}
In this paper, the fragmentation functions of both pions and kaons are determined by fitting the HERMES~\cite{HERMES:2012uyd} experimental multiplicity data with the extended Field-Feynman recursive model, and their uncertainties are also estimated by the Hessian method. Compared with the previous work~\cite{Hua:2003ie}, the SU(2) flavor symmetry breaking properties of $ud$ quarks of different meson fragmentation functions are discussed, and we show that the fragmentation functions of kaons have a bigger SU(2) flavor symmetry breaking effect of $ud$ quarks than these of pions. The results obtained are generally compatible with existing parametrizations of
fragmentation functions of pions and kaons.

\begin{acknowledgments}
This work is supported by National Natural Science Foundation of China under Grant No.~12075003.
\end{acknowledgments}

\appendix
\section{}\label{appendix_a}
It can be easily derived that the negatively charged meson fragmentation functions observe charge conjugation in the Field-Feynman model. With the help of Eq.~\eqref{mainFF}, we can get the fragmentation functions of the negatively charged pion:
\begin{align}
    D_{\bar{u}}^{\pi^-}(z) &=\beta_{ud}^{\pi} f_u^{\pi}(1-z)+\beta_{uu}^{\pi}\beta_{ud}^{\pi} \overline{F}_u^{\pi}(z), \label{Dpimu}\\
	D_{d}^{\pi^-}(z) &=\beta_{du}^{\pi} f_d^{\pi}(1-z)+\beta_{du}^{\pi}\beta_{dd}^{\pi} \overline{F}_d^{\pi}(z), \label{Dpimdbar}\\
	D_{\bar{d}}^{\pi^-}(z) &=\beta_{du}^{\pi}\beta_{dd}^{\pi} \overline{F}_d^{\pi}(z), \\
	D_{u}^{\pi^-}(z) &=\beta_{uu}^{\pi}\beta_{ud}^{\pi} \overline{F}_u^{\pi}(z),	\\
	D_{s}^{\pi^-}(z) &=D_{\bar{s}}^{\pi^-}(z)=\beta_{su}^{\pi}\beta_{sd}^{\pi} \overline{F}_s^{\pi}(z);
\end{align}
and these of the negatively charged kaon:
\begin{align}
    D_{\bar{u}}^{K^-}(z) &=\beta_{us}^{K} f_u^{K}(1-z)+\beta_{uu}^{K}\beta_{us}^{K} \overline{F}_u^{K}(z), \label{Dkmu}\\
	D_{s}^{K^-}(z) &=\beta_{su}^{K} f_s^{K}(1-z)+\beta_{su}^{K}\beta_{ss}^{K} \overline{F}_s^{K}(z), \label{Dkmsbar}\\
	D_{\bar{s}}^{K^-}(z) &=\beta_{su}^{K}\beta_{ss}^{K} \overline{F}_s^{K}(z), \\
	D_{u}^{K^-}(z) &=\beta_{uu}^{K}\beta_{us}^{K} \overline{F}_u^{K}(z),	\\
	D_{d}^{K^-}(z) &=D_{\bar{d}}^{K^-}(z)=\beta_{du}^{K}\beta_{ds}^{K} \overline{F}_d^{K}(z).
\end{align}
Obviously, there exist charge conjugation relations $D_q^{h^+}=D_{\bar{q}}^{h^-}$. But for the neutral meson, the relations are complex. With Eq.~\eqref{mainFF}, we can derive the neutral pion fragmentation functions
\begin{align}
    D^{\pi^0}_u(z) &=D^{\pi^0}_{\bar{u}}(z)=\frac{1}{2}\beta^{\pi}_{uu}f^{\pi}(1-z) + \frac{1}{2}\left[(\beta^{\pi}_{uu})^2+(\beta^{\pi}_{ud})^2\right]\overline{F}^{\pi}(z), \\
    D^{\pi^0}_d(z) &=D^{\pi^0}_{\bar{d}}(z)=\frac{1}{2}\beta^{\pi}_{dd}f^{\pi}(1-z) + \frac{1}{2}\left[(\beta^{\pi}_{dd})^2+(\beta^{\pi}_{du})^2\right]\overline{F}^{\pi}(z), \\
    D^{\pi^0}_s(z) &= D^{\pi^0}_{\bar{s}} = \frac{1}{2}\left[(\beta^{\pi}_{su})^2+(\beta^{\pi}_{sd})^2\right]\overline{F}^{\pi}(z).
\end{align}
It is shown that $D^{\pi^0}_q=D^{\pi^0}_{\bar{q}}$ since $\pi^0$ is the anti-particle of itself. But there is no relation \(D^{\pi^0}_q = \dfrac{1}{2}[D^{\pi^+}_q+D^{\pi^-}_q]\) since the SU(2) flavor symmetry breaking of $u$ and $d$ quarks. Unlike the neutral pion, there are two kinds of neutral kaons, $K^0$ and $\bar{K}^0$, with constituents $d\bar{s}$ and $\bar{d}s$. With the same steps, we can get the fragmentation functions of $K^0$:
\begin{align}
    D_{d}^{K^0}(z) &=\beta_{ds}^{K} f_d^{K}(1-z)+\beta_{dd}^{K}\beta_{ds}^{K} \overline{F}_d^{K}(z), \label{Dk0d}\\
	D_{\bar{s}}^{K^0}(z) &=\beta_{sd}^{K} f_s^{K}(1-z)+\beta_{sd}^{K}\beta_{ss}^{K} \overline{F}_s^{K}(z), \label{Dk0sbar}\\
	D_{s}^{K^0}(z) &=\beta_{sd}^{K}\beta_{ss}^{K} \overline{F}_s^{K}(z), \\
	D_{\bar{d}}^{K^0}(z) &=\beta_{dd}^{K}\beta_{ds}^{K} \overline{F}_d^{K}(z),	\\
	D_{u}^{K^0}(z) &=D_{\bar{u}}^{K^0}(z)=\beta_{ud}^{K}\beta_{us}^{K} \overline{F}_u^{K}(z);
\end{align}
and that of $\bar{K}^0$:
\begin{align}
    D_{\bar{d}}^{\bar{K}^0}(z) &=\beta_{ds}^{K} f_d^{K}(1-z)+\beta_{dd}^{K}\beta_{ds}^{K} \overline{F}_d^{K}(z), \label{Dk0bard}\\
	D_{s}^{\bar{K}^0}(z) &=\beta_{sd}^{K} f_s^{K}(1-z)+\beta_{sd}^{K}\beta_{ss}^{K} \overline{F}_s^{K}(z), \label{Dk0barsbar}\\
	D_{\bar{s}}^{\bar{K}^0}(z) &=\beta_{sd}^{K}\beta_{ss}^{K} \overline{F}_s^{K}(z), \\
	D_{d}^{\bar{K}^0}(z) &=\beta_{dd}^{K}\beta_{ds}^{K} \overline{F}_d^{K}(z),	\\
	D_{u}^{\bar{K}^0}(z) &=D_{\bar{u}}^{\bar{K}^0}(z)=\beta_{ud}^{K}\beta_{us}^{K} \overline{F}_u^{K}(z).
\end{align}
Apparently there exist relations $D^{K^0}_q=D^{\bar{K}^0}_{\bar{q}}$ since $K^0$ and $\bar{K}^0$ are anti-particles to each other. But there are also no relations $D^{K^+}_u = D^{K^0}_d$ and $D^{K^-}_{\bar{u}} = D^{\bar{K}^0}_{\bar{d}}$ due to the SU(2) flavor symmetry breaking of $u$ and $d$ quarks.

\section{}\label{appendix_b}
Because we made assumptions about the relationship between parameters $\beta^{\pi}_{su}$ and $\beta^{\pi}_{sd}$ for the pion in the fitting procedure (for the kaon, the parameters are $\beta^{K}_{du}$ and $\beta^K_{ds}$), the results obtained may not be physical. However, we can still get the value ranges of these parameters through some mathematical analysis.\par
For the pion, we assume that the relationship between these two parameters is $\beta^{\pi}_{su} = \beta^{\pi}_{sd}$. But in fact, the fitting of experimental data can only ensure that their product is certain. Combined with the conditions given by the model and their products determined in the fitting, we can get the following equations:
\begin{align}
    \beta^{\pi}_{su}\beta^{\pi}_{sd}=0.430^2&=0.1849,\\
    \beta^{\pi}_{su}+\beta^{\pi}_{sd}&<1.
\end{align}
By solving the equations, we get the range of two parameters: $0.245<\beta^{\pi}_{su} (\text{or }\beta^{\pi}_{sd}) <0.755$.\par
Similar to the pion, but the relationship between two parameters is $\beta^{K}_{ds}=1-2\beta^{K}_{du}$ for the kaon. Correspondingly, the equations are
\begin{align}
    \beta^{K}_{du}\beta^{K}_{ds} = 0.471\times 0.058 &=0.027318, \\
    \beta^{K}_{du}+\beta^{K}_{ds} &< 1.
\end{align}
Because the mass of $s$ quark is much larger than that of $u$ and $d$ quarks, there should be two additional constraints:
\begin{align}
\beta^K_{ds}&<\beta^K_{du}, \\
\beta^{K}_{ds}&<\beta^{K}_{dd}=1-\beta^K_{du}-\beta^K_{ds}.
\end{align}
Solving the equations with constraints, we get the range $0.058<\beta^{K}_{ds}<0.165<\beta^{K}_{du}(\text{or }\beta^{K}_{dd})<0.942$.
\bibliography{mybib}
\end{document}